# Mechanochemical Synthesis of the Lead-Free Double Perovskite $Cs_2[AgIn]Br_6$ and its Optical Properties.

Joachim Breternitz,[1]* Sergiu Levcenko,[1] Hannes Hempel,[1] Galina Gurieva,[1] Alexandra Franz,[1] Andreas Hoser,[1] and Susan Schorr[1,2]

*1 Helmholtz-Zentrum Berlin für Materialien und Energie, Structure and Dynamics of Energy Materials, Hahn-Meitner-Platz 1, 14109 Berlin, Germany.*

*2 Freie Universität Berlin, Department Geosciences, Malteserstraße 74-100, Berlin, Germany.*

## Abstract

Hitting hard on the binary halides yields in the formation of $Cs_2[AgIn]Br_6$. The lead-free double perovskite marks, although not usable itself, a further step forward in finding sustainable and durable perovskite materials for photovoltaic applications. $Cs_2[AgIn]Br_6$ is one of the prominent examples of double perovskites materials that have been suggested to circumvent the use of lead compounds in perovskite solar cells. We herein report the successful synthesis of the material using a mechanochemical approach. It crystallizes in an elpasolite-type structure, an ordered perovskite superstructure, with a cell parameter of $a$ = 11.00 Å. However, the compound exhibits a relatively large optical bandgap of 2.36 eV and is unstable under illumination, which impedes its use as solar absorber material at this early stage. Still, substitution of lead and the potential of this synthesis method are promising as well as the fruitful combination of theoretical considerations with experimental materials discovery.

Hybrid organic-inorganic perovskite solar cells are clearly one of the most intriguing developments within the field of energy materials in recent years.[1] Their quick development is, however, in stark contrast to some their inherent problems, which are most prominently the comparably low stability under working conditions as well as the toxicity of lead.[2] Additionally, lead is not only a scientific hurdle, for instance due to the complicated modelling of the organic cation orientation,[3] but also a legal one. The use of lead is strictly limited for electrical devices in most countries, for instance through the RoHS directive in the European Union,[4] and similar regulations in most countries. Therefore, considerable efforts have been invested in the development of lead-free and all-inorganic perovskite materials to overcome both the toxicity as well as the stability issue related to hybrid organic-inorganic lead halides.[5] An intriguing class of compounds that are bearing these properties are double perovskites of the general formula $A^{I}_2[B^{I}B'^{III}]X^{-I}_6$. In this material class the B position in $A^{I}B^{II}X^{-I}_3$ perovskites is substituted by equal amounts of single and triple charged cations. A number of stable double perovskites with B′ = Bi have been successfully synthesized,[6] but they generally have an indirect bandgap,[7] which gives them a considerable disadvantage in performance versus direct bandgap materials. Materials with B′ = In, on the other hand, have also been proposed through theoretical considerations,[7] but have proven to be more challenging to synthesize. $Cs_2[AgIn]Br_6$, for instance, was found to be thermodynamically stable but only in a very narrow region of the chemical potential.[7] Other calculations did actually predict an instability of the material compared to ternary compounds in the system.[8] It is hence expectable that the formation of this compound is challenging, and was in fact claimed unlikely to be achieved based on Goldschmidt tolerance factor considerations.[9] It should be noted that a number of recent publications report on the synthesis $Cs_2[AgIn]Cl_6$,[9,10] but its bromide analogue has, to the best of our knowledge, not been reported so far.

We used a mechanochemical approach that previously been used successfully for the synthesis of lead halide perovskite materials.[11-15] This technique is advantageous for the formation of compounds with narrow existence region.[16] First, the confined space of the reaction avoids loss of starting material and hence allows to control the final composition by the overall composition of the starting materials. Second, the reactions take place near ambient conditions and without any solvent. This prevents the formation of undesired phases through differences in solubility of the different starting materials or intermediate compounds as in a high temperature synthesis. Using a planetary ball mill with a stoichiometric mixture of the binary metal halides as starting materials, we obtained a bright yellow powder, which was treated in a nitrogen-filled glovebox due to its anticipated sensitivity towards moisture. Further details of synthesis and characterisation may be found in the supplementary information.

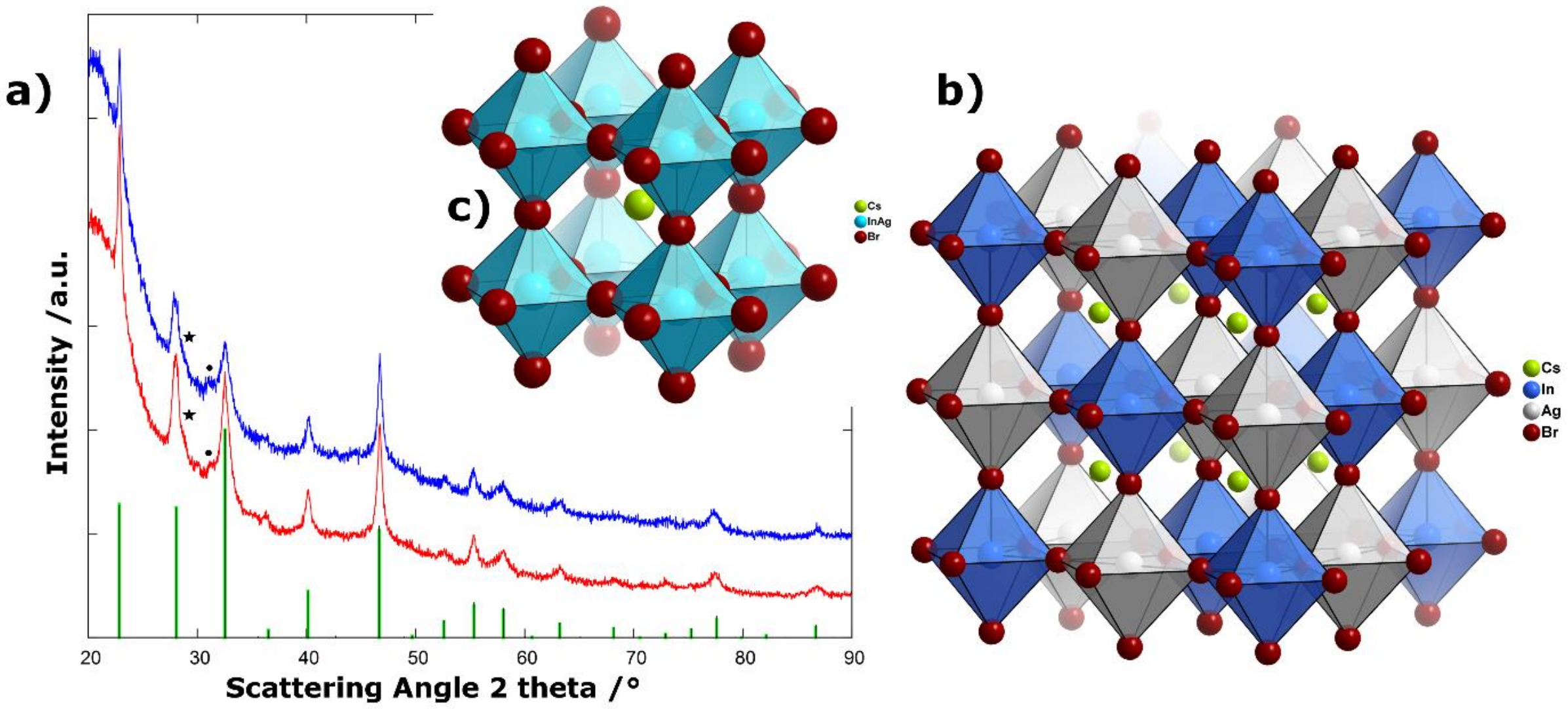

Figure 1. a) X-ray diffraction patterns of $Cs_2[AgIn]Br_6$ produced through milling times of 100 min (red) and 720 min (blue) (normalized and shifted for visibility) compared to the theoretical peak positions and intensities of the elpasolite-type structure (green sticks) and perovskite-type structure (blue sticks – largely overlapping the green sticks) The circles and stars mark residual reflections of AgBr and CsBr respectively. Structure representations of the ordered elpasolite-type structure (b) and disordered perovskite-type structure (c).

The X-ray diffraction pattern of the reaction product showed a pattern of mediocre quality (Figure 1a), which is mainly due to the small coherent domain size, as is characteristic for mechanochemical syntheses.[17] It should also be noted, that the synthesis of this compound using other methods proved difficult as it was reportedly not possible to obtain $Cs_2[AgIn]Br_6$ through solution synthesis[9] or solid state reaction.[18] Therefore, the disadvantages of a mechanochemical approach for further synthesis, namely the low crystallinity of the obtained product, are outweighed by the fact that this method is the only one so far to produce this compound. To evaluate the approximate crystallite size, we have used a Williamson-Hall type relationship[19] between the refined Lorentzian peak shape function and the crystallite size. This yielded in an approximate crystallite size of 10 nm (Figure S5). Attempts to anneal the product at temperatures up to 200 °C were unsuccessful and hence, an increase of the crystallinity could not be achieved. This was further confirmed in a temperature-dependent X-ray diffraction measurement, where the sample was heated under nitrogen gas flow (Figures S6-S8). It is evident that the double perovskite phase decomposes between 140 °C and 160

°C with a further phase change at 220 °C (Figure S7). Given the complex powder diffraction pattern, we were not successful to distinguish the exact composition of the decomposition product, yet. However, the powder pattern does neither match to the starting materials, nor to the phases identified as competing phases within the system.[7,8] Nevertheless, we were able to perform a Rietveld refinement of the compound in an elpasolite type structure (Figure 1b) using Jana2006,[20] which is also adopted by the analogous chloride compound $Cs_2[AgIn]Cl_6$.[9] The full details of the Rietveld refinement may be found in the ESI (Figure S1 and S2 and Table S1 – Table S14). Our refined cell parameter of $a$ = 10.997(5) Å is larger than the one found for $Cs_2[AgIn]Cl_6$ ($a$ = 10.47 Å),[9] which is in line with the differences in ionic radii for $Br^-$ ($r$ = 1.96 Å) and $Cl^-$ ($r$ = 1.81 Å).[21] In fact, the difference in cell parameters $\Delta a$ = 0.52 Å between the bromide compound reported here and the chloride compound synthesized by Volonakis *et al.*[9] coincides very closely with the difference in atomic radii (taking into account that two anions sit on each cell edge) $2\cdot\Delta d = 4\cdot\Delta r$ = 0.6 Å. Given the poor crystallinity of the powder pattern and the background introduced from the airtight sample holder, the Rietveld refinements yield relatively large uncertainties. The excellent agreement of our found cell parameters with the values expectable based on the chloride analogue therefore add another level of confidence in the findings presented herein. In order to reach a stable refinement, the isotropic displacement parameters of $In^{3+}$ and $Ag^+$ were constrained to a common value, since both ions are isoelectronic and hence indiscernible from X-ray diffraction.

The elpasolite-type structure can be thought of as a perovskite-type derived structure, where the octahedral, corner sharing, B-sites are alternatingly occupied by Ag and In atoms. Corner-sharing $[InBr_6]$ and $[AgBr_6]$ octahedra form the backbone, in the voids of which the Cs atoms occupy dodecahedrally coordinated positions. Since $In^{3+}$ and $Ag^+$ are indiscernible from X-ray diffraction, the question whether the *B*-cations are ordered alternatingly in the elpasolite-type structure, or disordered in the perovskite-type structure (with $2\cdot a_{\mathrm{Perovskite}} = a_{\mathrm{elpasolite}}$) cannot be answered directly from X-ray diffraction (Figure 1a-c, see ESI for Rietveld refinements in both structural models). While the differences in neutron scattering lengths ($b_{\mathrm{Ag}}$ = 5.922 fm, $b_{\mathrm{In}}$ = 4.065–0.0539i fm) should, in principle, suffice to make a clear distinction between the ordered and disordered structural model, the low crystallinity of our samples and the consequently large peak widths, prevented any conclusion on the ordering of the structure with this method (Figure S3 and S4).

Nevertheless, we are confident to attribute the structure to the ordered elpasolite type, due to another characteristic of this structure, as compared to the perovskite type structure. While the bromide positions in the undistorted perovskite-type are fixed by symmetry to be exactly in the middle between two metal positions, and hence only allow one *M*–Br distance, no such restriction exists in the elpasolite-type structure, were the Br atom can be distorted from the symmetrical position between two edge-sharing octahedra. This supplementary degree of freedom allows the formation of two distinct sorts of $[MBr_6]$ octahedra with different M–Br distances. When refining this site, the Br atom shifts away from the symmetrical position at ¼ towards one of the metal sites yielding in a larger octahedron with $d_{\mathrm{M–Br}}$ = 2.84(5) Å and a smaller octahedron with $d_{\mathrm{M–Br}}$ = 2.66(5) Å. Given the comparably large difference in ionic radii between $Ag^+$ ($r$ = 1.15 Å) and $In^{3+}$ ($r$ = 0.8 Å) in octahedral coordination,[21] it is only logic to assume an ordered elpasolite-type structure with $Ag^+$ in the larger octahedra and $In^{3+}$ in the smaller octahedra. Applying the bond-valence concept to these compounds ($R_{\mathrm{In-Br}}$ = 2.41, $R_{\mathrm{Ag-Br}}$ = 2.22 and $b$ = 0.37 taken from Brese and O'Keeffe[22]), the valence sums of In and Ag in these octahedra of 3.05 and 1.12 are close to their charges. In other words, the expected bond lengths from trivalent Indium $d_{\mathrm{In-Br}}$ = 2.666 Å and monovalent silver $d_{\mathrm{Ag-Br}}$ = 2.883 Å are very close to the values observed in the structure. In fact, the very same shift is observed in $Cs_2AgInCl_6$, where the distances are $d_{\mathrm{Ag-Cl}}$ = 2.85 Å and $d_{\mathrm{In-Cl}}$ = 2.38 Å.[9] The situation for $Cs_2AgBiBr_6$ is less clear as the differences in size between the Bi-Br and Ag-Br octahedra are less distinct. As a matter of fact, two of the three structures reported in the ICSD report the Ag-Br distance to be smaller than the Bi-Br distance,[23,24] while the third reports the inverse situation.[6] The second case appears more physical at first sight, since the Shannon radius for $Bi^{3+}$ (1.03 Å) is smaller than the one for $Ag^{+}$.

To further confirm the physicality of this anion shifting, we performed a structure optimization of the elpasolite-type structure. Firstly, we note that the cell parameters of the optimized structure are slightly larger than the observed cell parameters (Table S14). Using the PBE exchange-correlation potential often produces slightly overestimated cell parameters,[25] but the optimized cell parameters do correlate very well with the ones prior obtained using more expensive calculation methods.[7,8] Herein, we observe a significant shift of the Br atoms away from the symmetrical position towards the In atoms. The M-Br distances in the optimized structure of $d_{\text{Ag-Br}}$ = 2.869 Å and $d_{\text{In-Br}}$ = 2.738 Å are slightly longer than our observed values, which is mainly due to the fact that the optimized unit cell parameters are approximately 2 % larger than the observed ones. However, the increase is nearly isotropic and the ratio $d_{\text{Ag-Br}}/d_{\text{In-Br}}$ (DFT) = 1.048 is close to the observed one ($d_{\text{Ag-Br}}/d_{\text{In-Br}}$ = 1.068).

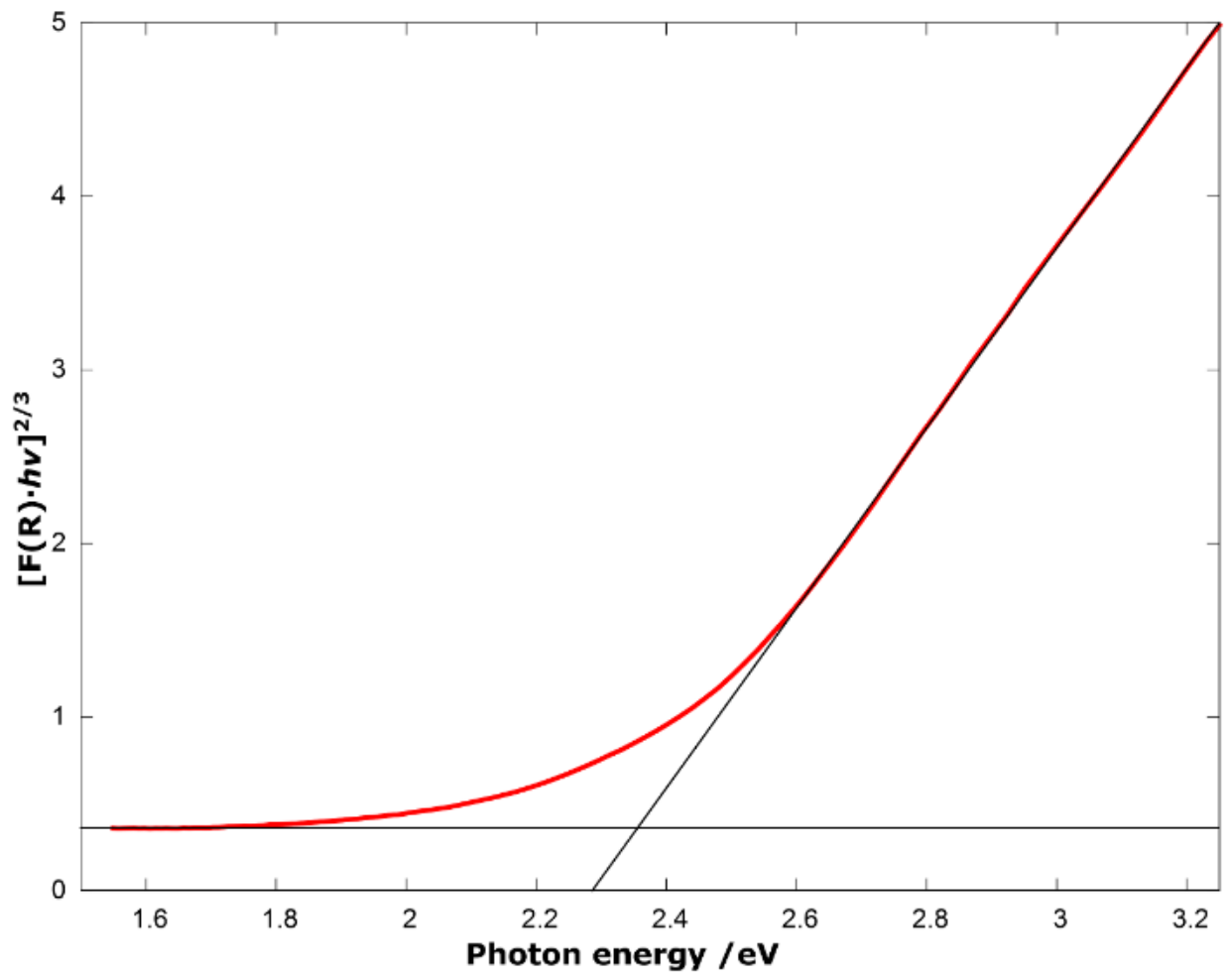

Figure 2: Tauc-plot for a direct forbidden transition (red line) with a linear fit in the region of 2.6-3.25 eV (black line, *f*(x) = 5.187*x-11.858). The horizotal black line give the base line as *g*(x) = 0.36.

While our finding confirms the general conclusion of prior theoretical work that the compound should be stable, we find an optical bandgap of $E_g$ = 2.36 eV from diffuse reflectance measurements (**Figure 2**). The optical bandgap was obtained by treating the diffuse reflectance data with the Kubelka-Munk function $F(R) = (1-R)^2/2R$ (where $R$ is the reflectance of the sample) to obtain a calculated absorption combined with a Tauc-plot treatment, as successfully applied on halide semiconductors previously.[26] Plotting the data as for a direct forbidden transition with $[F(R)\cdot hv]^{2/3}$ gives a very good linear fit over a wide region above the optical bandgap. This optical bandgap is significantly higher than predicted from theoretical calculations ($E_g$ in the range of 1.33 – 1.5 eV).[7,8] It is remarkable that the same effect is observed for the chloride analogue, where the bandgap error is $E_g$(expt.)-$E_g$(theor.) = 0.9 eV.[7-9] In a recent theoretical study, this behavior was explained by parity-forbidden transitions that make excitations at the bandgap level unlikely.[27] When zooming on the onset of the Tauc function below the optical bandgap, one finds an increase of the absorption at about 1.7 eV and hence much closer to the theoretically calculated bandgap (Figure S10). This discrepancy may be taken as a hint for a parity-forbidden transition in this material as predicted by Mitzi *et al.*[27]

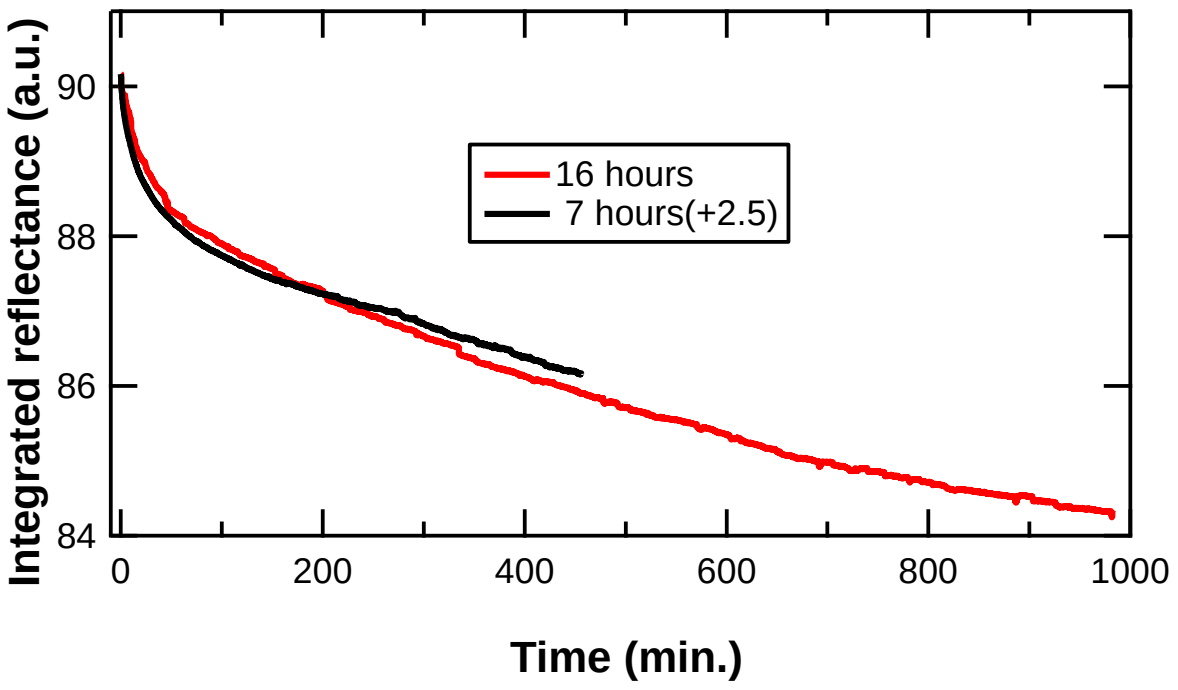

Figure 3: Integrated reflectance signals over the 400- 1100 nm range vs light exposure time. The measurement over 7 hours (red curve) is shifted by 2.5 in intensity for a better comparison with the measurement over 16 hours (black curve).

Not only that the observed optical bandgap of 2.36 eV is probably too large for direct use of this material in a solar cell, we further observed another problematic characteristic that will prevent this material from being used as solar absorber. When exposing the compound to light, a rather quick degradation of the material can be observed, clearly visible through a darkening of the illuminated area (Figure S11). To quantify the effect of light on $Cs_2[AgIn]Br_6$ compound, we performed long–term spectrally resolved reflectance measurements with a broadband light source on the sealed sample. The decay of the reflection below the band gap of $Cs_2[AgIn]Br_6$ (2.3 eV ≈ 540 nm) shows that the compound degrades quickly under light illumination (Figures 3, S12 - S14). One can quantify that approximately half of the relative change happening in the first ≈40 min (Figure 3). While the illuminated area is clearly darkened after the light exposure, the rest of the sample preserved its color during the measurement, showing that the degradation was caused by the light, rather than being an intrinsic effect of the material (Figure S13). The stability in dark conditions is also supported by prior diffuse reflectance measurements of freshly prepared material and after 2 hours of storage in in the dark (Figure S9). The light-induced decomposition of the material might happen in analogy to AgBr, which decomposes into elemental silver nanoparticles and $Br_{2(g)}$ under light illumination.[28] A similar effect of light on the analogous Bi-containing double perovskite $Cs_2AgBiBr_6$, but not with the same severity, has been observed.[6,24] Further studies on nanocrystals of the material showed the formation of silver nanoparticles during the degradation.[29] We would expect that the decomposition process in $Cs_2[AgIn]Br_6$ would work through a similar pathway. The degradation of the material results in a notable change of the reflectance in the region below the bandgap (500-900 nm) (Figure S14-S17). Furthermore, we observed that the irradiated spot appeared less dark after a period of time (Figure S13), which could rather be a consequence of an overall sample degradation than a "healing effect" but a more detailed study would clearly be beneficial to understand the light degradation of this material.

In conclusion, we present the successful mechanochemical synthesis of $Cs_2[AgIn]Br_6$ which is of great interest due to its predicted suitability as potential solar absorber material. However, the measured optical bandgap of the compound is 2.36 eV and is approx. 1 eV larger than predicted by theory. This relatively large band gap and the observation of a light-induced degradation are significant challenges for the application of $Cs_2[AgIn]Br_6$ in photovoltaics should be addressed in further studies. Still, the proof of its existence in this work is a powerful demonstration of the possibilities of *in silico* materials discovery and their targeted synthesis to screen the chemical space in the most efficient way.

**Acknowledgements**

The authors would like to acknowledge Helmholtz-Zentrum Berlin für Materialien und Energie for the use of E6 and E9 neutron instruments at BERII and the use of the X-ray corelab for ambient and non-ambient XRD measurements. The authors further acknowledge Pascal Becker for the assistance in the reflection measurements and René Gunder for his assistance with the non-ambient temperature measurements.

## REFERENCES

[1] a) Kojima, A.; Teshima, K.; Shirai, Y.; Miyasaka, T., *J. Amer. Chem. Soc.* **2009**, *131*, 6050-6051; b) Green, M. A.; Ho-Baillie, A.; Snaith, H.J. *Nat. Photonics* **2014**, *8*, 506-514.

[2] *e.g.* M. Grätzel, *Nat. Mater.* **2014**, *13*, 838.

[3] a) A. Franz, D. M. Többens, S. Schorr, *Cryst. Res. Technol.* **2016**, *51*, 534; b) G. Schuck, D. M. Többens, M. Koch-Müller, I. Efthimopoulos, S. Schorr, *J. Phys. Chem. C* **2018**, *122*, 5227.

[4] Restriction of the use of certain hazardous substances Directive 2011/65/EU.

[5] M. H. Kumar, S. Dharani, W. L. Leong, P. P. Boix, R. R. Prabhakar, T. Baikie, C. Shi, H. Ding, R. Ramesh, M. Asta, M. Grätzel, S. G. Mhaisalkar, N. Mathews, *Adv. Mater.* **2014**, *26*, 7122.

[6] E. T. McClure, M. R. Ball, W. Windl, P. M. Woodward, *Chem. Mater.* **2016**, *28*, 1348.

[7] a) X.-G. Zhao, D. Yang, Y. Sun, T. Li, L. Zhang, L. Yu, A. Zunger, *J. Am. Chem. Soc.* **2017**, *139*, 6718; b) J. Xu, J.-B. Liu, B.-X. Liu, B. Huang, *J. Phys. Chem. Lett.* **2017**, *8*, 4391.

[8] Z. Xiao, K.-Z. Du, W. Meng, D. B. Mitzi, Y. Yan, *Angew. Chem. Int. Ed.* **2017**, *56*, 12107; *Angew. Chem.* **2017**,*129*, 12275.

[9] G. Volonakis, A. A. Haghighirad, R. L. Milot, W. H. Sio, M. R. Filip, B. Wenger, M. B. Johnston, L. M. Herz, H. J. Snaith, F. Giustino, *J. Phys. Chem. Lett.* **2017**, *8*, 772.

[10] a) J. Luo, S. Li, H. Wu, Y. Zhou, Y. Li, J. Liu, K. Li, F. Yi, G. Niu, J. Tang, *ACS Photonics* **2018**, *5*, 398; b) F. Locardi, M. Cirignano, D. Baranov, Z. Dang, M. Prato, F. Drago, M. Ferretti, V. Pinchetti, M. Fanciulli, S. Brovelli, L. De Trizio, L. Manna, *J. Amer. Chem. Soc.* **2018**, *140*, 12989.

[11] D. Prochowicz, M. Franckevičius, A. M. Cieślak, S. Zakeeruddin, M. Grätzel, J. Lewiński, *J. Mater. Chem. A* **2015**, *3*, 20772.

[12] A. D. Jodlowski, A. Yépez, R. Luque, L. Camacho, G. de Miguel, *Angew. Chem. Int. Ed.* **2016**, *55*, 14972; *Angew. Chem.* **2016**, *128*, 15196.

[13] Z.-Y. Zhu, Q.-Q. Yang, L.-F. Gao, L. Zhang, A.-Y. Shi, C.-L. Sun, Q. Wang, H.-L. Zhang, *J. Phys. Chem. Lett.* **2017**, *8*, 1610.

[14] D. Prochowicz and P. Yadav and M. Saliba and M. Saski and S. M. Zakeeruddin and J. Lewiński and M. Grätzel, *Sustainable Energy Fuels* **2017**, *1*, 689.

[15] P. Sadhukhan, S. Kundu, A. Roy, A. Ray, P. Maji, H. Dutta, S. K. Pradhan, S. Das, *Cryst. Growth Des.* **2018**, *18*, 3428.

[16] C. Suryanarayana, *Prog. Mater. Sci.* **2001,** *46*, 1.

[17] H.-J. Fecht, *Nanostruct. Mater.* **1995**, *6*, 33.

[18] K.-Z. Du, W. Meng, X. Wang, Y. Yan, D. B. Mitzi, *Angew. Chem. Int. Ed.* **2017**, *56*, 8158; *Angew. Chem.* **2017**, *129*, 8270.

[19] a) G. K. Williamson, W. H. Hall, *Acta Metall.* **1953**, *1*, 22; b) P. Barns, S. Jacques, M. Vickers, http://pd.chem.ucl.ac.uk/pdnn/peaks/sizedet.htm, accessed 03 August 2018.

[20] V. Petříček, M. Dušek, L. Palatinus, *Z. Kristallogr. Cryst. Mater.* **2014**, *229*, 345.

[21] R. D. Shannon, *Acta Crystallogr. A* **1976**, *32*, 751.

[22] N. E. Brese, K. O'Keeffe, *Acta Crystallogr. B* **1991**, *47*, 192.

[23] M. R. Filip, S. Hillman, A. A. Haghighirad, H. J. Snaith, F. Giustino, *J. Phys. Chem. Lett.* **2016**, *7*, 2579.

[24] A. H. Slavney, T. Hu, A. M. Lindenberg, H. I. Karunadasa, *J. Am. Chem. Soc.* **2016**, *138*, 2138.

[25] W.-J. Yin, J.-H. Yang, J. Kang, Y. Yan, S.-H. Wei, *J. Mater. Chem. A* **2015**, *3*, 8926.

[26] K. S. Virdi, Y. Kauffmann, C. Ziegler, P. Ganter, P. Blaha, B. V. Lotsch, W. D. Kaplan, C. Scheu, *J. Phys. Chem. C* **2016**, *120*, 11170.

[27] W. Meng, X. Wang, Z. Xiao, J. Wang, D. B. Mitzi, Y. Yan, *J. Phys. Chem. Lett.* **2017**, *8*, 2999.

[28] A. F. Hollemann, E. Wiberg, N. Wiberg, *Lehrbuch der Anorganischen Chemie, 101. Auflage*, de Gruyter, Berlin, Germany **1995**, 1394.

[29] Y. Bekenstein, J. C. Dahl, J. Huang, W. T. Osowiecki, J. K. Swabeck, E. M. Chan, P. Yang, A. P. Alivisatos, *Nano Lett.* **2018**, *18*, 3502.

**Supplementary Information**

## 1) Materials and Methods

*Synthesis*

Stoichiometric amounts of $InBr_3$ (Roth, 99.999 %), AgBr (Roth, 99.999 %) and CsBr (Alfa Aesar, 99 %) were used as supplied and weighed in a nitrogen filled glovebox ($O_2$ < 10 ppm, $H_2O$ < 10 ppm) to make up 2 g of reaction mixture and tightly sealed in a 45 ml ball-mill steel jar with 10 mm steel balls. The mechanochemical syntheses were performed in a Fritsch Pulverisette 7 premium line planetary ball mill with a ball-to-powder ratio of 1:16. Milling was performed in sequences of 10 min milling at 400 rpm, interrupted by 2 min of resting time and a sequence of 10 cycles. Increasing the reaction time to 72 cycles (10 min milling at 400 rpm and 5 min rest) produces a very similar result (Figure 1a) and further characterizations were performed on the material produced at shorter reaction time. It should be noted that all samples contained traces of the starting materials, but they were so small that the fractions could not be refined in the Rietveld refinements. We attribute their presence to the comparably inaccurate weighing procedure within the glovebox that is, for instance, affected by small changes in the pressure within the glovebox. More accurate weighing could for instance be obtained by increasing the overall mass of the reaction mixture in order to reduce the relative error of the scale.

*X-ray Diffraction*

X-ray diffraction patterns were recorded on a Bruker d8 Advance system with an air-tight sample dome, filled in the glovebox (see above), with Bragg-Brentano geometry. The diffractometer is equipped with a LynxEye detector and Ni-filtered Cu-$K_\alpha$ radiation (λ = 1.5418 Å). Rietveld Refinements of the data were performed with Jana2006,[S1] using a manual background and a supplementary 5-term Legendre polynomial function to account for the complicated background form mainly caused by the sample dome. The isotropic displacement parameters of Ag1 and In1 were constrained to a common value to account for their similarity in scattering. Very small reflections of the starting materials AgBr and CsBr can be found in the pattern, but their intensity is so low that a stable refinement could not be achieved, and they were hence excluded from the refinement. Based on the intensity of the side phases, we estimate their remaining amount to be below 1 vol-%. The temperature-dependent X-ray diffraction was recorded using a nitrogen flushed Anton Paar HTK1200N hotstage. A short contact of the sample with air of ≈10 sec could not be avoided during the sample loading, but was reduced to a minimum by filling the sample holder in the glovebox and transporting it to the instrument in a closed vessel. The reaction chamber was flushed with nitrogen prior to loading the sample. After being loaded, the sample was heated from 40 °C to 360 °C in steps of 20 K. At each step, a powder pattern was recorded for approx. 1 hour. Structure visualizations were performed using Diamond 3,[S2] and powder pattern simulations were made using PowderCell 2.4.[S3]

*Neutron diffraction*

A sample was measured on both the E9 instrument[S4] and the E6 instrument at the BERII research reactor in Berlin. The sample was measured at room temperature within an air-tight vanadium container (6 mm diameter) that was filled in the glovebox.

Measurement details E9: The sample was measured for 9 h with a static detector (λ = 1.7981Å, focussing in an angular range, where the 511 reflection of the elpasolite-type structure is located.

This reflection is not overlapping with a reflection from the perovskite-type and should hence clearly indicate the ordered elpasolite-type if observed.

Measurement details E6: The sample was measured in the complete angular range overnight (≈15 h) in order to cover as many reflections as possible with λ = 2.4 Å. The longer wavelength spreads the reflections further apart in angular space and hence helps to separate weak reflections. This has to be traded against stronger instrumental peak broadening compared to E9.

*DFT calculations*

GGA-DFT calculations were performed using Abinit 8.6.3[S5] with PBE exchange-correlation potential[S6] and pseudopotentials created with the ONCVPSP code.[S7] Abinit input files were created using the cif2cell program.[S8]

*Optical measurements*

Diffuse reflectance UV-VIS spectrometry was recorded using a Perkin-Elmer Lambda 950 spectrometer. To this end, the sample was mounted on the back of the integrating Ulbrichts-sphere and illuminated with monochromatic light from a tungsten-halogen lamp. The reflection light was measured by an avalanche photo diode and calibrated to a white standard. To protect the sample from moisture, it was prepared between two glass slides and sealed with high-viscosity vacuum grease and duct tape. Long term reflectometry measurements under constant illumination were conducted in a custom-built setup, which consist of the Thorlabs SLS2001 tungsten halogen broad band light source (center wavelength ≈ 1000 nm) and AvaSpec-2048XL high speed UV- and NIR sensitivity back-thinned CCD Spectrometer.

## 2) Rietveld Refinement of $Cs_2[AgIn]Br_6$

a) Refinement in Elpasolite-type structure

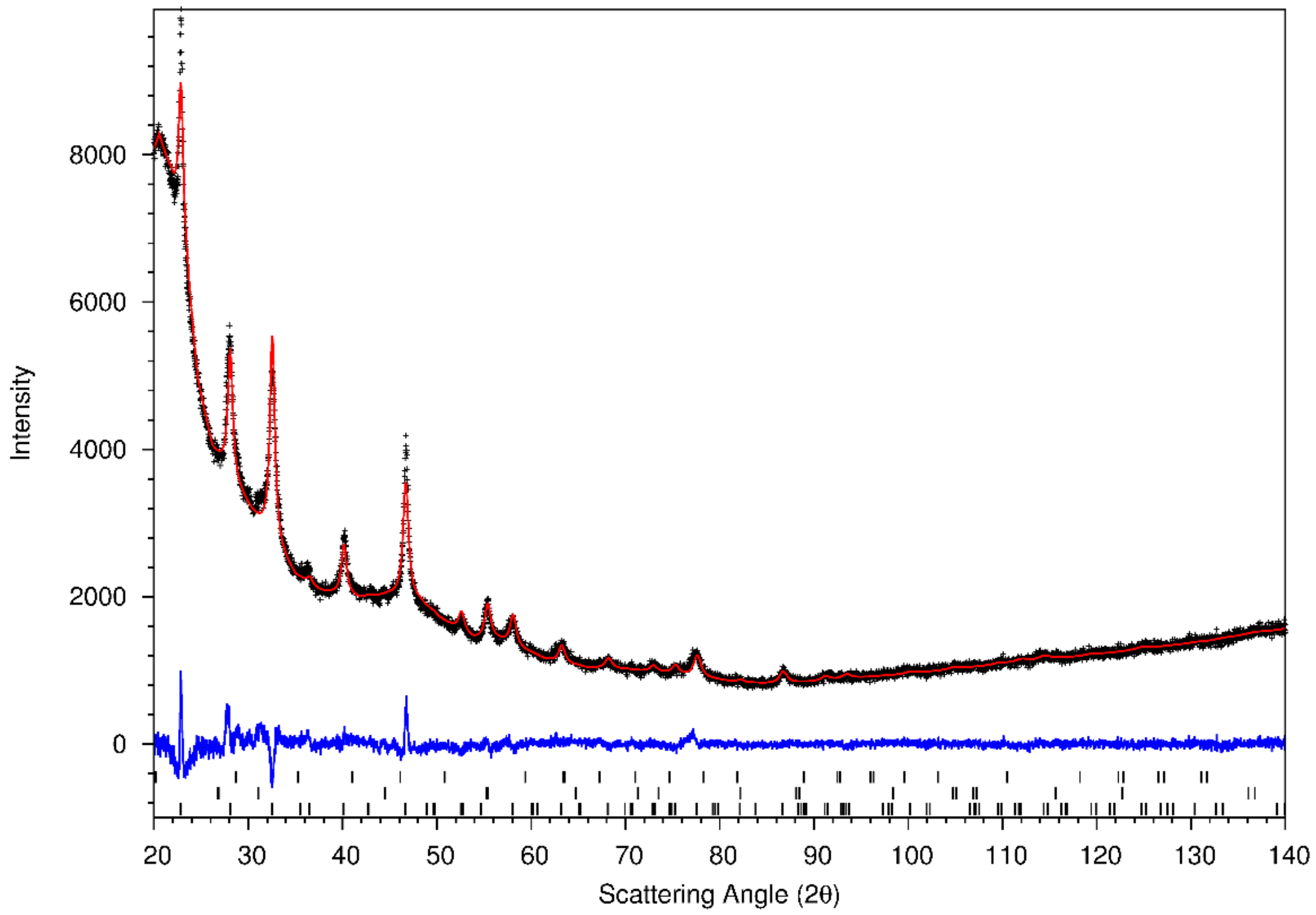

Figure S1: Observed (black crosses) and refined (red line) powder profile and their difference (blue line) together with the calculated peak positions (black ticks, bottom to top) for $Cs_2[AgIn]Br_6$, AgBr and CsBr. The latter two are shown for information but were not included in the refinement.

Table S1: Crystal data

| | |
|---|---|
| $AgBr_6Cs_2In$ | *F*(000) = 90 |
| $M_r$ = 967.9 | $D_x$ = 4.834 Mg m$^{-3}$ |
| Cubic, $Fm\bar{3}m$ | Cu *K*α radiation, λ = 1.5418 Å |
| Hall symbol: -F 4 2 3 | *T* = 293 K |
| *a* = 10.997 (5) Å | Particle morphology: irregular, visual examination |
| *V* = 1329.9 (10) Å$^3$ | yellow |
| *Z* = 4 | 10 × 10 mm |

Table S2: Data collection

| | |
|---|---|
| Bruker d8 advance diffractometer | Data collection mode: reflection |
| Radiation source: sealed tube, Bruker d8 advance | Scan method: continuous |
| Ni filter monochromator | $2\theta_{min}$ = 20.02°, $2\theta_{max}$ = 139.97° |
| Specimen mounting: powder on off-cut Si with dome | |

Table S3: Refinement

| | |
|---|---|
| $R_p$ = 0.028 | 16 parameters |
| $R_{wp}$ = 0.036 | 0 restraints |
| $R_{exp}$ = 0.024 | 4 constraints |
| *R*(*F*) = 0.095 | Weighting scheme based on measured s.u.'s |
| 5360 data points | $(\Delta/\sigma)_{max}$ = 0.007 |
| Excluded region(s): from 19.98 to 20.000, from 139.995 to 140.135 | Background function: Manual background combined with 5 Legendre polynoms |
| Profile function: Pseudo-Voigt | Preferred orientation correction: none |

Table S4: Fractional atomic coordinates and isotropic or equivalent isotropic displacement parameters (Å$^2$) for (I)

| | *x* | *y* | *z* | $U_{iso}$*/$U_{eq}$ |
|---|---|---|---|---|
| Cs1 | 0.25 | 0.25 | 0.25 | 0.038 (11)* |
| In1 | 0 | 0 | 0 | 0.065 (12)* |
| Ag1 | 0.5 | 0.5 | 0.5 | 0.065 (12)* |
| Br1 | 0.242 (4) | 0 | 0 | 0.056 (10)* |

Table S5: Geometric parameters (Å, º) for (I)

| | | | |
|---|---|---|---|
| In1—Br1 | 2.66 (5) | Ag1—Br1$^{vi}$ | 2.84 (5) |
| | | | |
| Br1—In1—Br1$^{i}$ | 180.0 (5) | Br1$^{vi}$—Ag1—Br1$^{viii}$ | 90 |
| Br1—In1—Br1$^{ii}$ | 90 | Br1$^{vi}$—Ag1—Br1$^{ix}$ | 90 |
| Br1—In1—Br1$^{iii}$ | 90 | Br1$^{vi}$—Ag1—Br1$^{x}$ | 90 |
| Br1—In1—Br1$^{iv}$ | 90 | Br1$^{vi}$—Ag1—Br1$^{xi}$ | 90 |
| Br1—In1—Br1$^{v}$ | 90 | Br1$^{vii}$—Ag1—Br1$^{viii}$ | 90 |
| Br1$^{i}$—In1—Br1$^{ii}$ | 90 | Br1$^{vii}$—Ag1—Br1$^{ix}$ | 90 |
| Br1$^{i}$—In1—Br1$^{iii}$ | 90 | Br1$^{vii}$—Ag1—Br1$^{x}$ | 90 |
| Br1$^{i}$—In1—Br1$^{iv}$ | 90 | Br1$^{vii}$—Ag1—Br1$^{xi}$ | 90 |
| Br1$^{i}$—In1—Br1$^{v}$ | 90 | Br1$^{viii}$—Ag1—Br1$^{ix}$ | 180.0 (5) |
| Br1$^{ii}$—In1—Br1$^{iii}$ | 180.0 (5) | Br1$^{viii}$—Ag1—Br1$^{x}$ | 90 |
| Br1$^{ii}$—In1—Br1$^{iv}$ | 90 | Br1$^{viii}$—Ag1—Br1$^{xi}$ | 90 |
| Br1$^{ii}$—In1—Br1$^{v}$ | 90 | Br1$^{ix}$—Ag1—Br1$^{x}$ | 90 |
| Br1$^{iii}$—In1—Br1$^{iv}$ | 90 | Br1$^{ix}$—Ag1—Br1$^{xi}$ | 90 |
| Br1$^{iii}$—In1—Br1$^{v}$ | 90 | Br1$^{x}$—Ag1—Br1$^{xi}$ | 180.0 (5) |
| Br1$^{iv}$—In1—Br1$^{v}$ | 180.0 (5) | In1—Br1—Ag1$^{xii}$ | 180.0 (5) |
| Br1$^{vi}$—Ag1—Br1$^{vii}$ | 180.0 (5) | | |

Symmetry codes: (i) -*x*, -*y*, *z*; (ii) *z*, *x*, *y*; (iii) *z*, -*x*, -*y*; (iv) *y*, *z*, *x*; (v) -*y*, *z*, -*x*; (vi) *x*, *y*+1/2, *z*+1/2; (vii) -*x*+1, -*y*+1/2, *z*+1/2; (viii) *z*+1/2, *x*, *y*+1/2; (ix) *z*+1/2, -*x*+1, -*y*+1/2; (x) *y*+1/2, *z*+1/2, *x*; (xi) -*y*+1/2, *z*+1/2, -*x*+1; (xii) *x*, *y*-1/2, *z*-1/2.

b) Refinement in perovskite-type structure

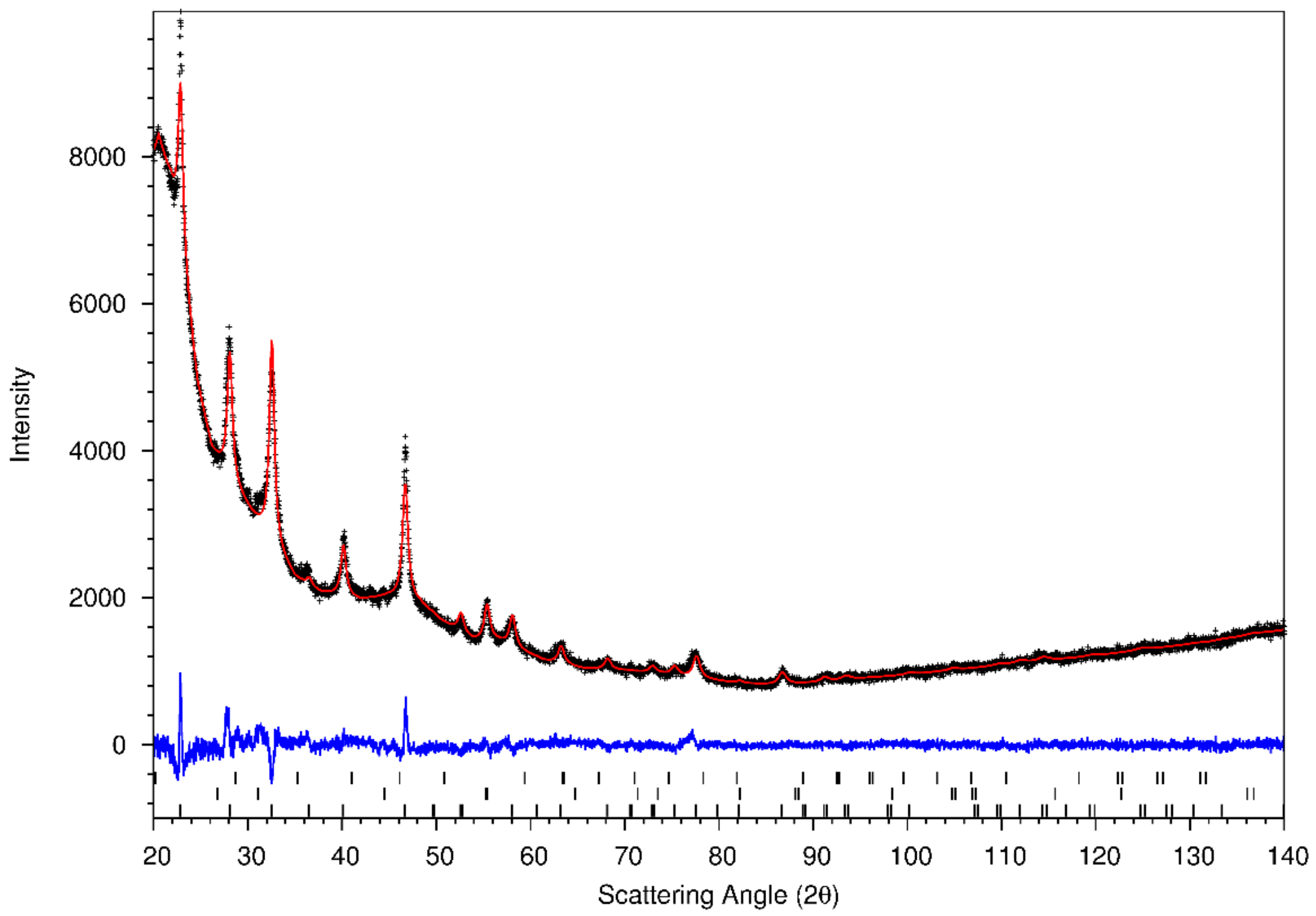

Figure F2: Observed (black crosses) and refined (red line) powder profile and their difference (blue line) together with the calculated peak positions (black ticks, bottom to top) for $Cs_2[AgIn]Br_6$, AgBr and CsBr. The latter two are shown for information but were not included in the refinement.

Table S6: Crystal data

| | |
|---|---|
| $Ag_{0.5}Br_3CsIn_{0.5}$ | $F(000) = 90$ |
| $M_r = 484$ | $D_x = 4.833$ Mg m$^{-3}$ |
| Cubic, $Pm\bar{3}m$ | Cu $K\alpha$ radiation, $\lambda = 1.5418$ Å |
| Hall symbol: -P 4 2 3 | $T = 293$ K |
| $a = 5.4991$ (18) Å | Particle morphology: irregular, visual examination |
| $V = 166.29$ (9) Å$^3$ | yellow |
| $Z = 1$ | 10 × 10 mm |

Table S7: Data collection

| | |
|---|---|
| Bruker d8 advance diffractometer | Data collection mode: reflection |
| Radiation source: sealed tube, Bruker d8 advance | Scan method: continuous |
| Ni filter monochromator | $2\theta_{min} = 20.02°$, $2\theta_{max} = 139.97°$ |
| Specimen mounting: powder on off-cut Si with dome | |

Table S8: Refinement

| | |
|---|---|
| $R_p$ = 0.027 | 15 parameters |
| $R_{wp}$ = 0.035 | 0 restraints |
| $R_{exp}$ = 0.024 | 4 constraints |
| $R(F)$ = 0.090 | Weighting scheme based on measured s.u.'s |
| 5360 data points | $(\Delta/\sigma)_{max}$ = 0.018 |
| Excluded region(s): from 19.98 to 20.000, from 139.995 to 140.135 | Background function: Manual background combined with 5 Legendre polynoms |
| Profile function: Pseudo-Voigt | Preferred orientation correction: none |

Table S9: Fractional atomic coordinates and isotropic or equivalent isotropic displacement parameters (Å$^2$) for (I)

| | *x* | *y* | *z* | $U_{iso}$*/$U_{eq}$ | Occ. (<1) |
|---|---|---|---|---|---|
| Cs1 | 0.5 | 0.5 | 0.5 | 0.046 (9)* | |
| In1 | 0 | 0 | 0 | 0.071 (10)* | 0.5 |
| Ag1 | 0 | 0 | 0 | 0.071 (10)* | 0.5 |
| Br1 | 0.5 | 0 | 0 | 0.066 (8)* | |

Table S10: Geometric parameters (Å, º) for (I)

| | | | |
|---|---|---|---|
| In1—Ag1 | 0 | Ag1—Br1$^{i}$ | 2.7496 (18) |
| | | | |
| Ag1—In1—Br1$^{iii}$ | 0 | Br1$^{i}$—Ag1—Br1 | 180.0 (5) |
| Ag1—In1—Br1$^{iv}$ | 0 | Br1$^{i}$—Ag1—Br1$^{ii}$ | 90 |
| Ag1—In1—Br1$^{v}$ | 0 | Br1$^{i}$—Ag1—Br1$^{iii}$ | 90 |
| Br1$^{i}$—In1—Br1 | 180.0 (5) | Br1$^{i}$—Ag1—Br1$^{iv}$ | 90 |
| Br1$^{i}$—In1—Br1$^{ii}$ | 90 | Br1$^{i}$—Ag1—Br1$^{v}$ | 90 |
| Br1$^{i}$—In1—Br1$^{iii}$ | 90 | Br1—Ag1—Br1$^{ii}$ | 90 |
| Br1$^{i}$—In1—Br1$^{iv}$ | 90 | Br1—Ag1—Br1$^{iii}$ | 90 |
| Br1$^{i}$—In1—Br1$^{v}$ | 90 | Br1—Ag1—Br1$^{iv}$ | 90 |
| Br1—In1—Br1$^{ii}$ | 90 | Br1—Ag1—Br1$^{v}$ | 90 |
| Br1—In1—Br1$^{iii}$ | 90 | Br1$^{ii}$—Ag1—Br1$^{iii}$ | 180.0 (5) |
| Br1—In1—Br1$^{iv}$ | 90 | Br1$^{ii}$—Ag1—Br1$^{iv}$ | 90 |
| Br1—In1—Br1$^{v}$ | 90 | Br1$^{ii}$—Ag1—Br1$^{v}$ | 90 |
| Br1$^{ii}$—In1—Br1$^{iii}$ | 180.0 (5) | Br1$^{iii}$—Ag1—Br1$^{iv}$ | 90 |
| Br1$^{ii}$—In1—Br1$^{iv}$ | 90 | Br1$^{iii}$—Ag1—Br1$^{v}$ | 90 |
| Br1$^{ii}$—In1—Br1$^{v}$ | 90 | Br1$^{iv}$—Ag1—Br1$^{v}$ | 180.0 (5) |
| Br1$^{iii}$—In1—Br1$^{iv}$ | 90 | In1—Br1—In1$^{vi}$ | 180.0 (5) |
| $Br1^{iii}$—In1—$Br1^{v}$ | 90 | In1—Br1—Ag1 | 0.0 (5) |
| $Br1^{iv}$—In1—$Br1^{v}$ | 180.0 (5) | In1—Br1—$Ag1^{vi}$ | 180.0 (5) |
| In1—Ag1—$Br1^{i}$ | 0 | $In1^{vi}$—Br1—Ag1 | 180.0 (5) |
| In1—Ag1—Br1 | 0 | $In1^{vi}$—Br1—$Ag1^{vi}$ | 0.0 (5) |
| In1—Ag1—$Br1^{ii}$ | 0 | Ag1—Br1—$Ag1^{vi}$ | 180.0 (5) |

Symmetry codes: (i) *x*-1, *y*, *z*; (ii) *z*, *x*-1, *y*; (iii) *z*, *x*, *y*; (iv) *y*, *z*, *x*-1; (v) *y*, *z*, *x*; (vi) *x*+1, *y*, *z*.

### 3) Neutron Measurements

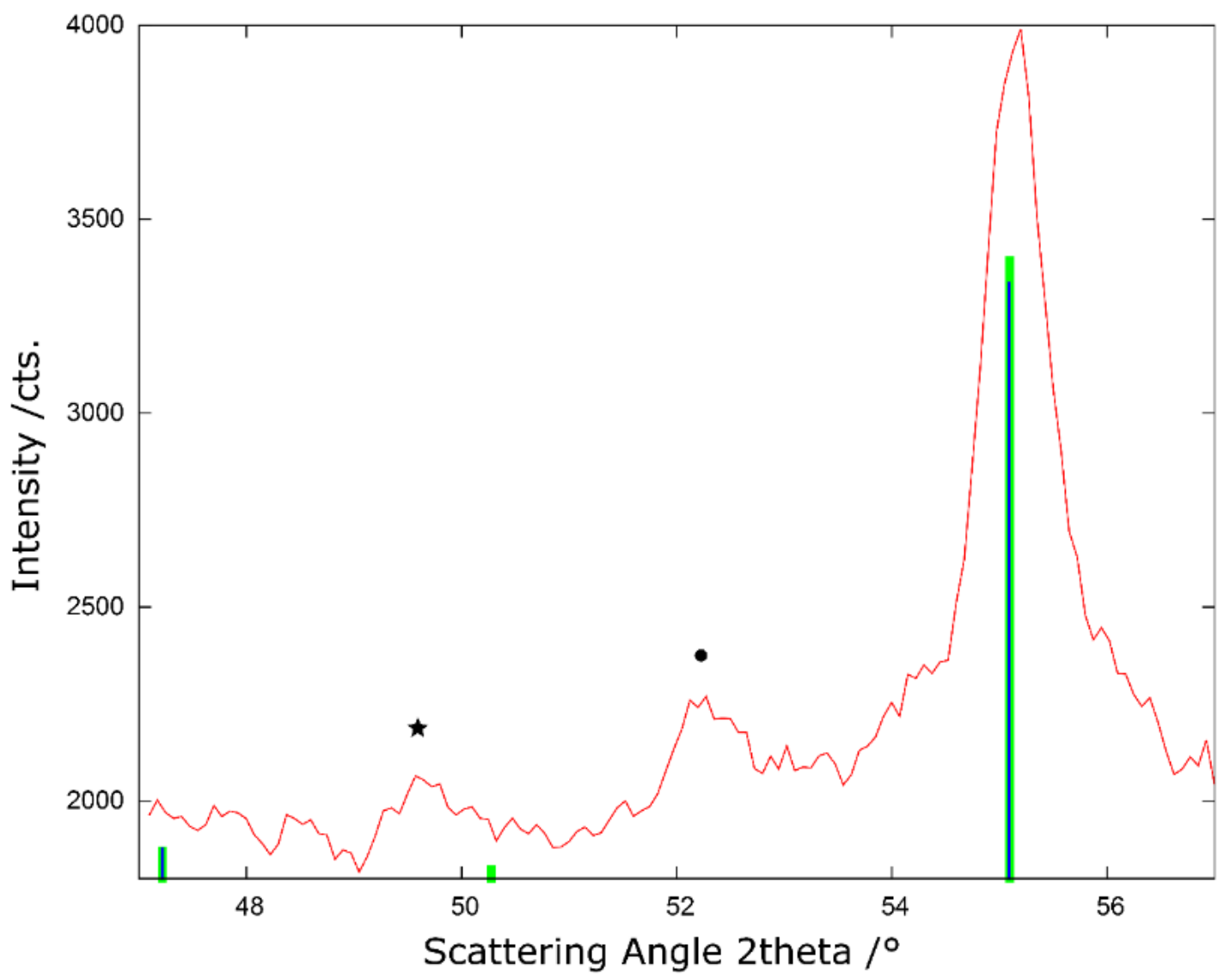

Figure S3: Measured diffractogram of $Cs_2[AgIn]Br_6$ (red line) at E9 with the calculated peak positions and intensities for the perovskite-type structure model (blue sticks) and elpasolite-type model (green sticks). The reflections from unreacted AgBr and CsBr are marked with a circle and star respectively.

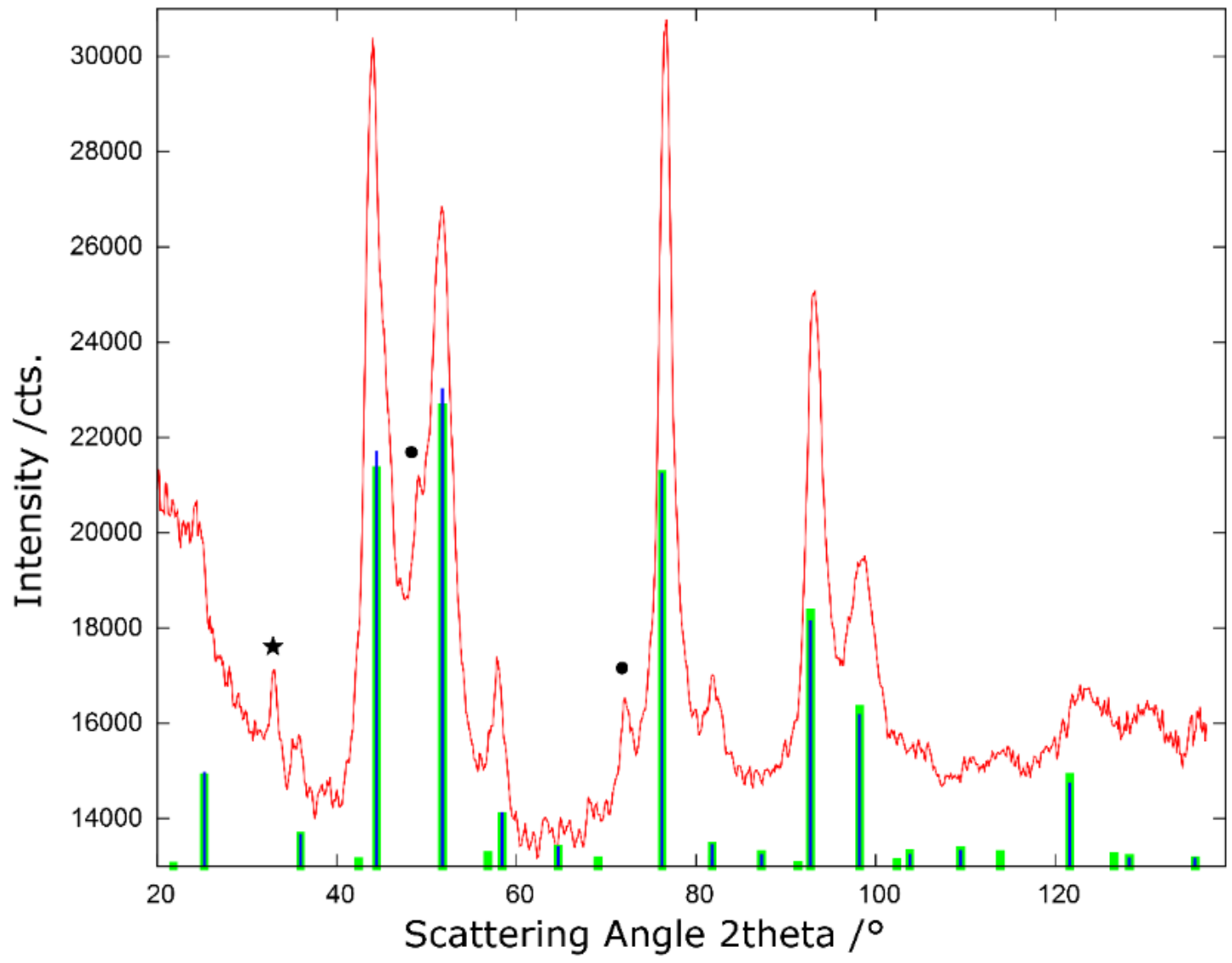

Figure S4: Measured diffractogram of $Cs_2[AgIn]Br_6$ (red line) at E6 with the calculated peak positions and intensities for the perovskite-type structure model (blue sticks) and elapsolite-type model (green sticks). The reflections from unreacted AgBr and CsBr are marked with a circle and star respectively.

## 4) Crystallite Size Analysis

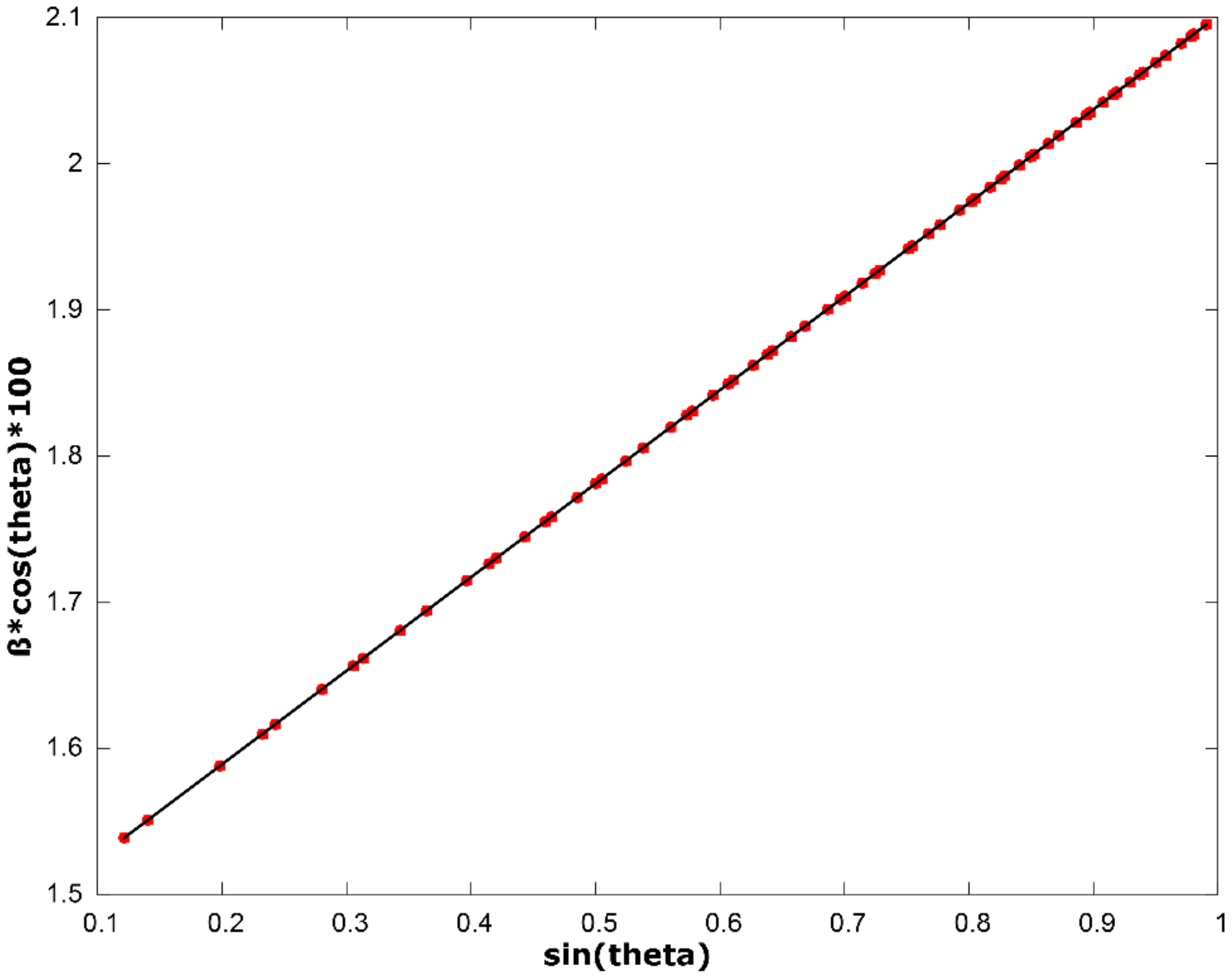

Figure S5: Williamson-Hall plot for the line broadening in $Cs_2[AgIn]Br_6$ refined in the elpasolite-type structure (see section 1 in the SI) as obtained from the Lorentzian peak profile. The data points were fitted with a linear function as $f(x) = 0.6402*x + 1.4610$.

The peak shape function was analysed using Jana2006 using its Williamson-Hall analysis tool. Herein, the peak broadening over the angular range (as defined through the Lorentzian terms used in the Rietveld refinement) was plotted against sin(theta). Given the strong broadening of the reflections and the comparably rough estimate of the method, we assumed the instrumental broadening to be negligible. With the line broadening term *ß* as the intersection of the linear fit with the y-axis (*ß* = 1.4609), the apparent crystallite size *L* can be calculated as $ß=K\cdot\lambda/L$, where K is a shape factor and is assumed to be 0.9-1.

For *K* = 0.9: *L* = 9.5 nm; For *K* = 1: *L* = 10.5 nm.

Given the rough estimate nature of this analysis, this absolute value should, nonetheless, not be overestimated.

## 4) Non-ambient XRD Measurements

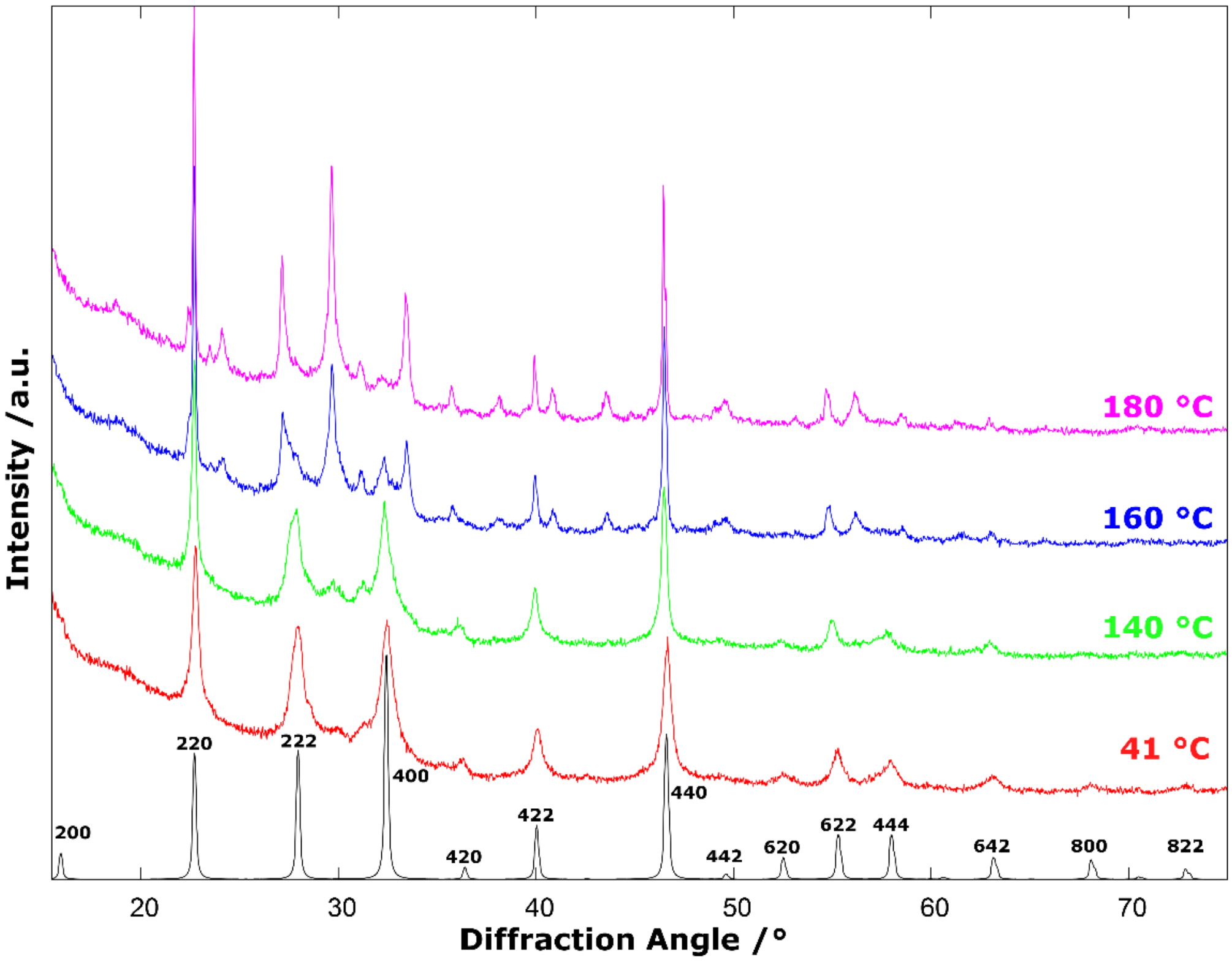

Figure S6: comparison of the XRD patterns obtained at 41 °C (red), 140 °C (green), 160 °C (blue) and 180 °C (pink) with the simulated powder pattern for $Cs_2[AgIn]Br_6$ as refined in the elpasolite-type structure. The hkl values for the major reflections are given and the diffraction patterns were shifted for better visibility.

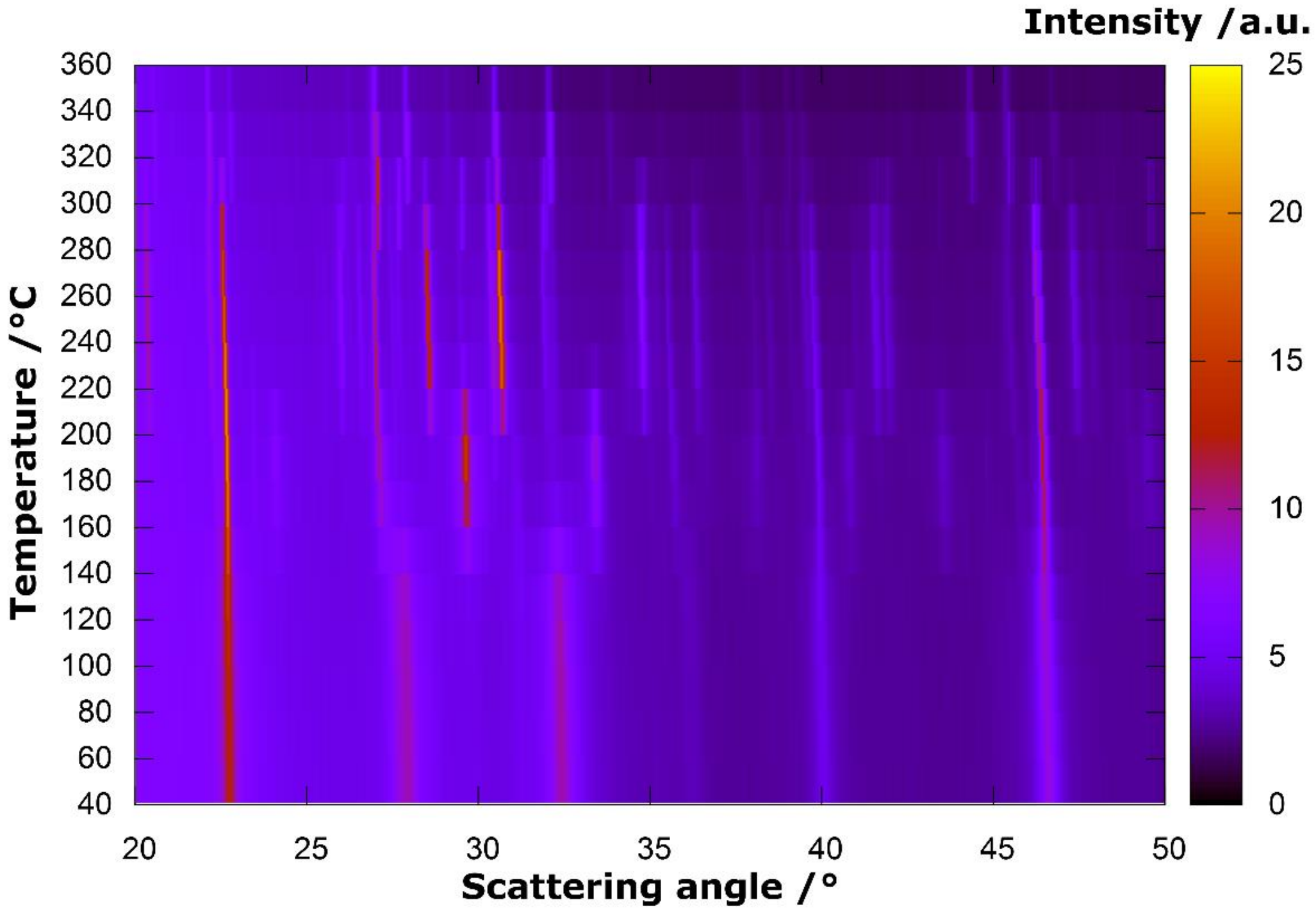

Figure S7: Heat map plot of the in-situ XRD measurements at higher temperatures. The tics on the y-axis signify the measurement temperature (for technical reasons, the lowest temperature measurement had to be run at 41 °C instead of 40 °C).

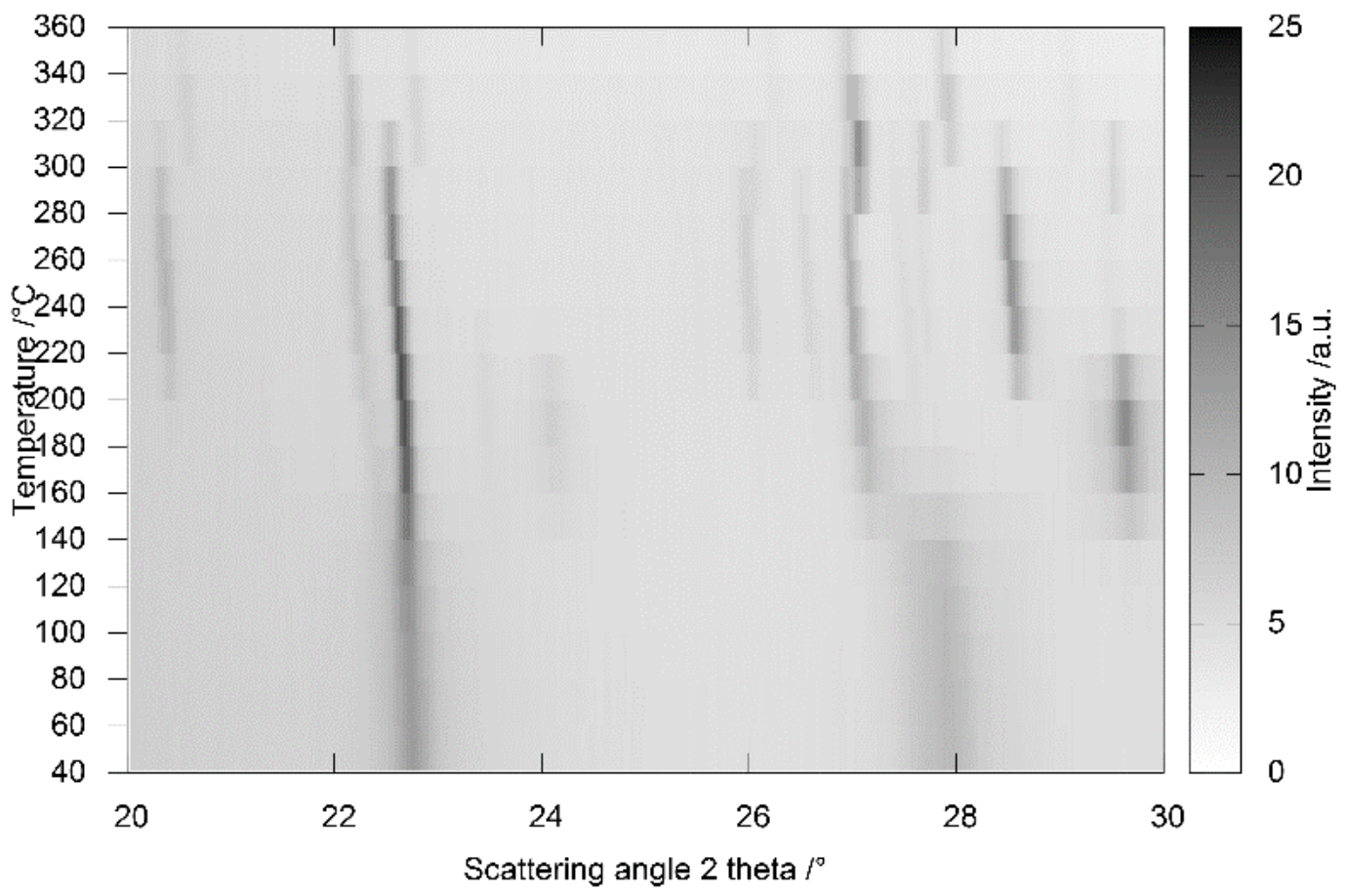

Figure S8: Film strip plot of the in-situ XRD measurements at higher temperatures. Zoom on the region between 20° and 30° 2 theta.

## 5) DFT calculations

Cut-off energy and k-point grind were converged first until $\Delta E_{tot}$ < 0.0001 Ht (2.7 meV). The convergence criterion for all further calculations was set to $\Delta E_{tot} < 10^{-10}$ Ht.

Table S11: Cutoff Energy convergence. The selected parameter is highlighted.

| Cutoff Energy /Ht | Total Energy $E_{tot}$ /Ht | ($E_{tot+1}$)-$E_{tot}$ /mHt |
|---|---|---|
| 35 | -363.044493 | -1.47219 |
| **40** | **-363.0459652** | **-0.08306** |
| 45 | -363.0460482 | -0.01966 |
| 50 | -363.0460679 | -0.02116 |
| 55 | -363.0460891 | |

Table S12: *k*-point grid convergence. The selected parameter is highlighted.

| k-point grid | Total Energy $E_{tot}$ /Ht | ($E_{tot+1}$)-$E_{tot}$ /mHt |
|---|---|---|
| 1 1 1 | -363.0372655 | -8.78279 |
| **2 2 2** | **-363.0460483** | **-0.06965** |
| 4 4 4 | -363.0461179 | -0.00067 |
| 6 6 6 | -363.0461186 | |

Structure optimisations were performed using the cell parameters as obtained from the refinement (a = 10.997 Å) as well as fully optimised ones. The latter is

Table S13: Structure optimisation results in fixed cell size.

| | *x* | *y* | *z* |
|---|---|---|---|
| Cs1 | 0.25 | 0.25 | 0.25 |
| In1 | 0 | 0 | 0 |
| Ag1 | 0.5 | 0.5 | 0.5 |
| Br1 | 0.245867 | 0 | 0 |
| | | | |
| *d*(Ag-Br) /Å | 2.795 | | |
| *d*(In-Br) /Å | 2.704 | | |

Table S14: Structure optimisation with optimised cell parameters.

| | | | |
|---|---|---|---|
| *a* /Å | 11.21228 | | |
| | | | |
| | *x* | *y* | *z* |
| Cs1 | 0.25 | 0.25 | 0.25 |
| In1 | 0 | 0 | 0 |
| Ag1 | 0.5 | 0.5 | 0.5 |
| Br1 | 0.244159 | 0 | 0 |
| | | | |
| *d*(Ag-Br) /Å | 2.869 | | |
| *d*(In-Br) /Å | 2.738 | | |

## 6) Diffuse reflectance measurements

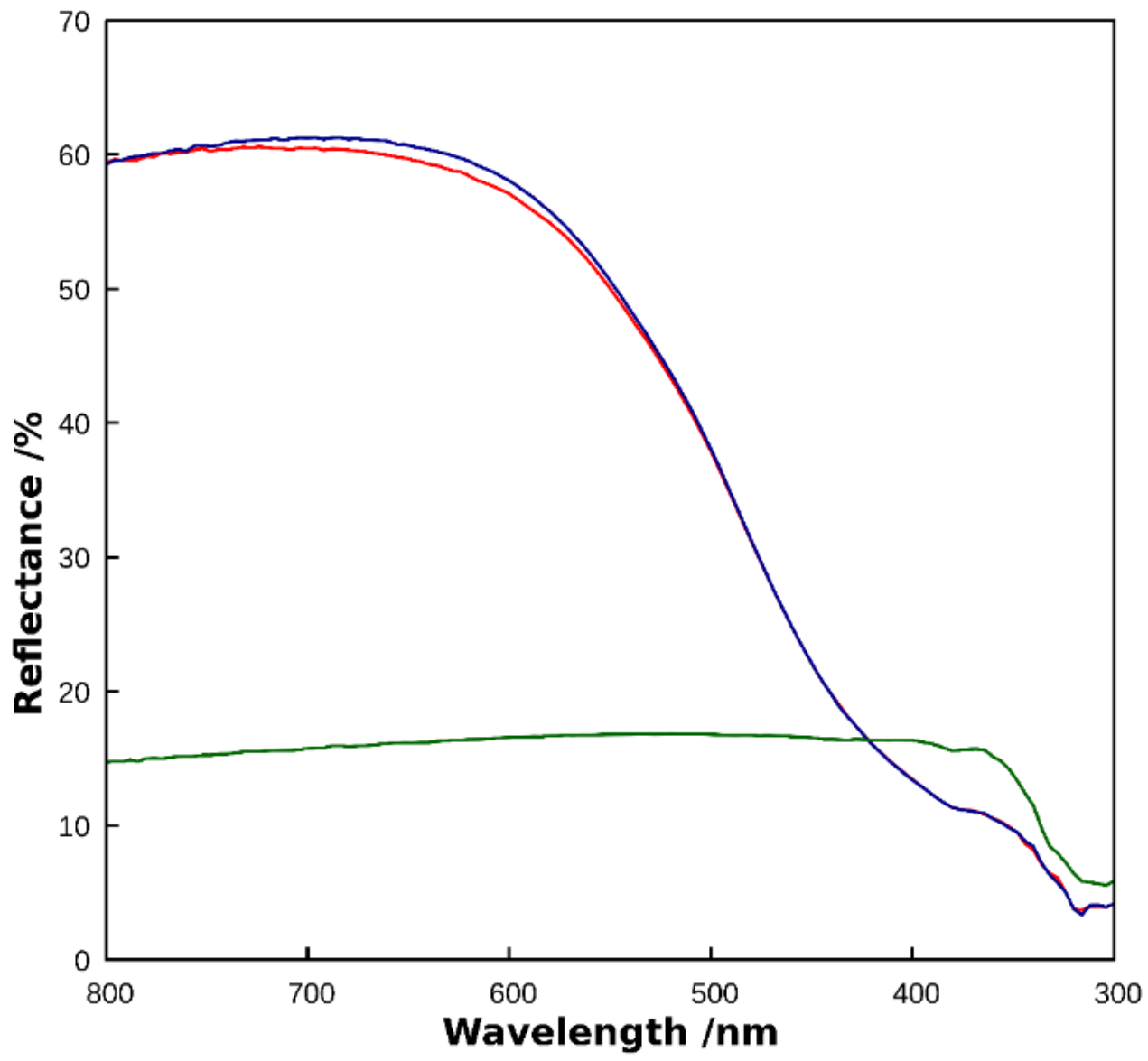

Figure S9: Diffuse reflectance measurement of $Cs_2[AgIn]Br_6$ as prepared (red line) as well as after two hours in the instrument (blue line). The green curve is a measurement of the pure glass used as sample holder. The absorption below 350 nm is caused by the glass.

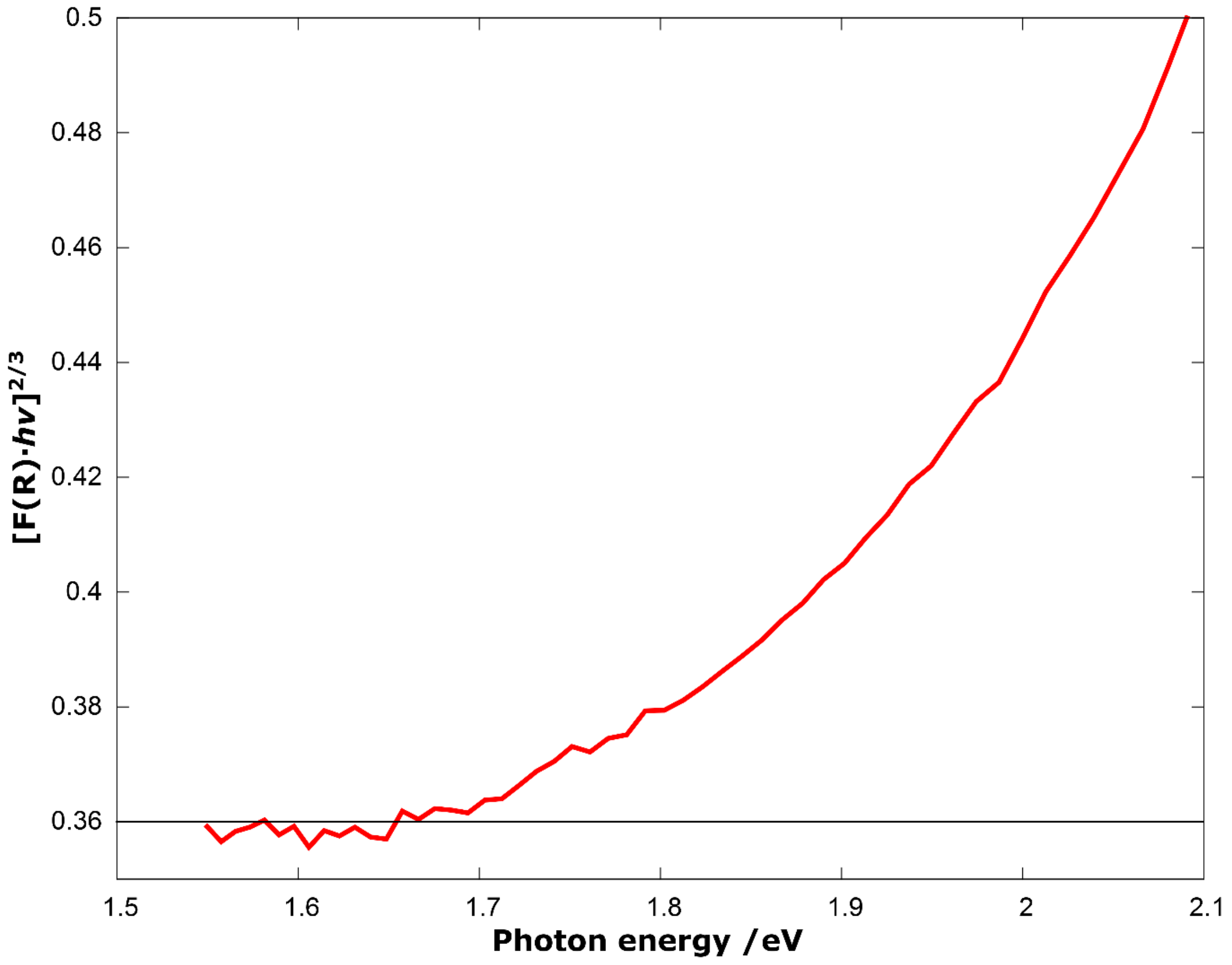

Figure S10: Tauc-plot for a direct forbidden transition close to the onset of light absorption (red line). The black line gives the horizontal base line as g(x) = 0.36.

**7) Long-time reflectance study**

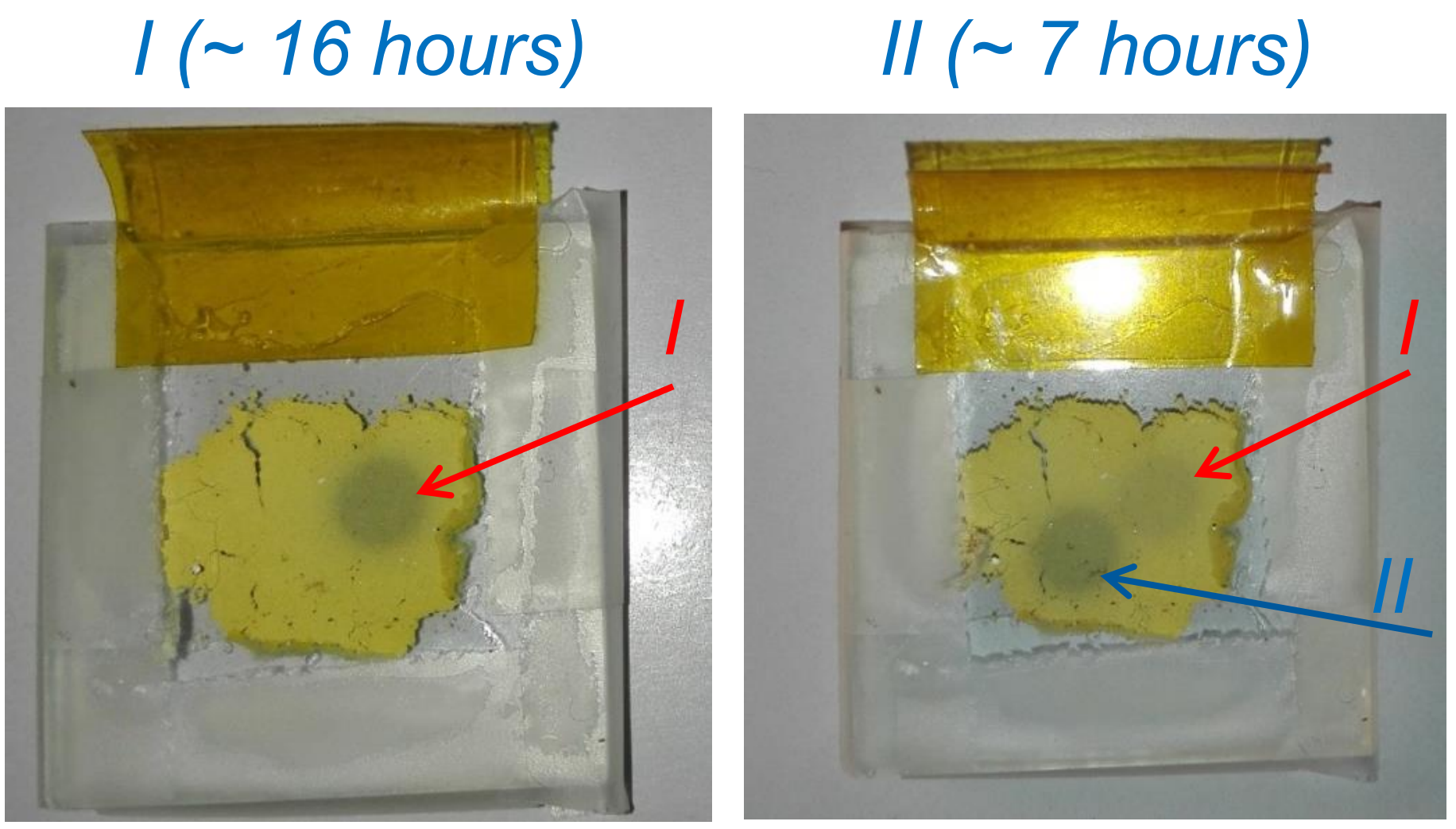

Figure S11: Photographs of the sample used for long-time reflectance testing after the first test of approximately 16 hours (left) and the second test of 7 hours. The illuminated spots are marked with red and blue arrows respectively.

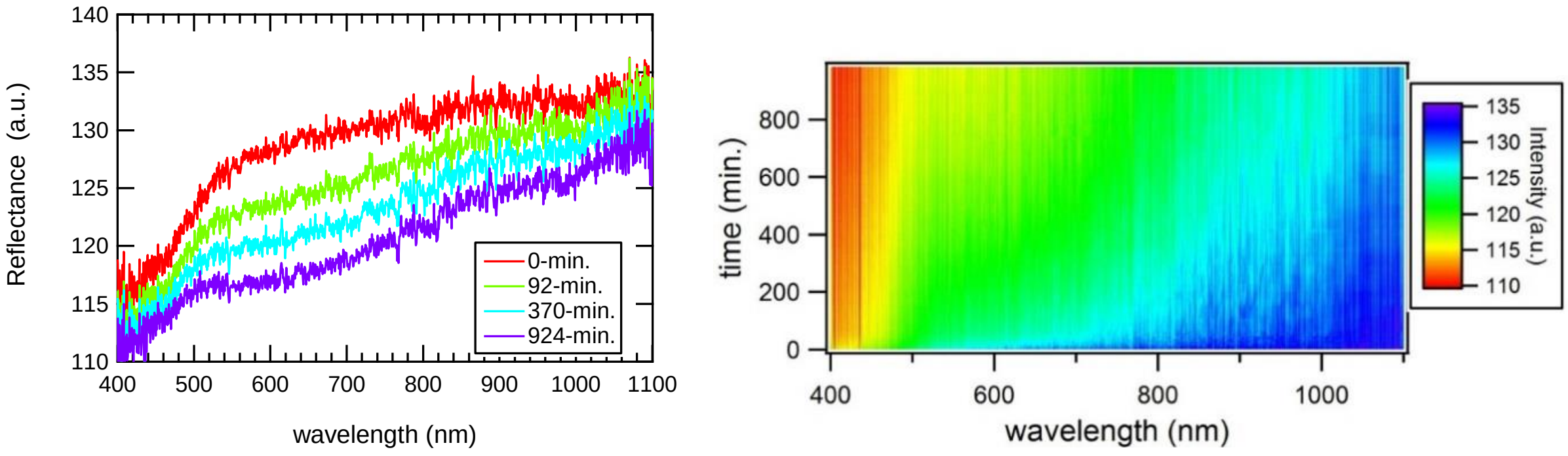

Figure S12: Reflectance spectra during measurement 1 (16 h) at given times (left) and as heatmap (right). The spectra clearly demonstrate the degradation of the sample over time.

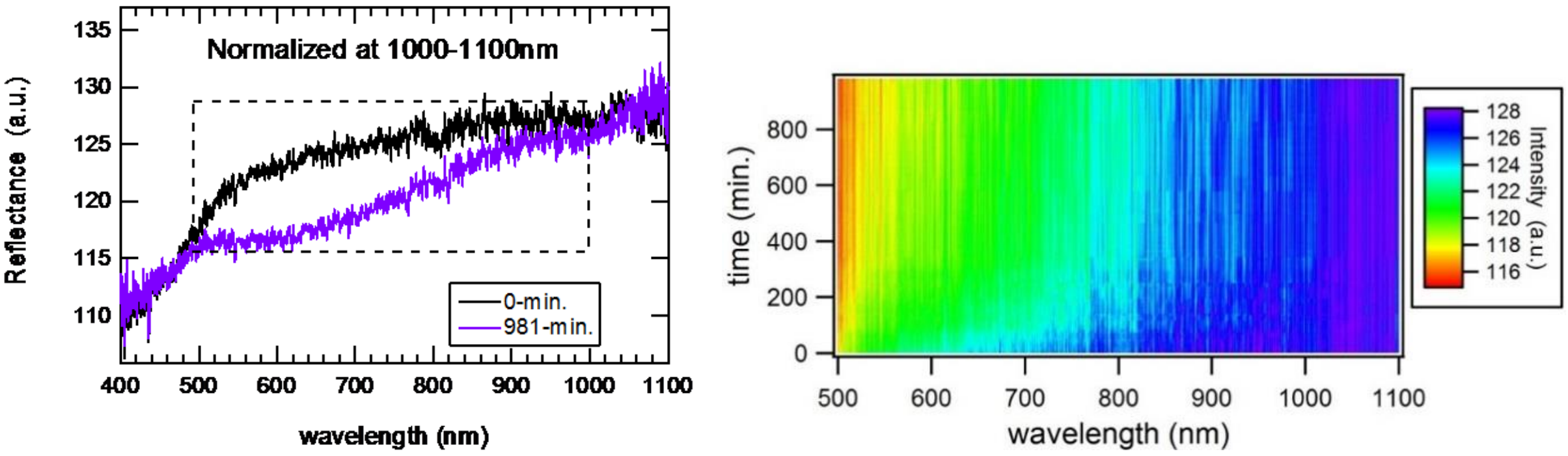

Figure S13: Reflectance spectra during measurement 1 (16 h) at given times (left) and as heatmap (right) normalised to the region 1000-1100 nm. Further to the overall loss in intensity, this demonstrates a stronger decrease in the region 500-900 nm than in the overall spectra.

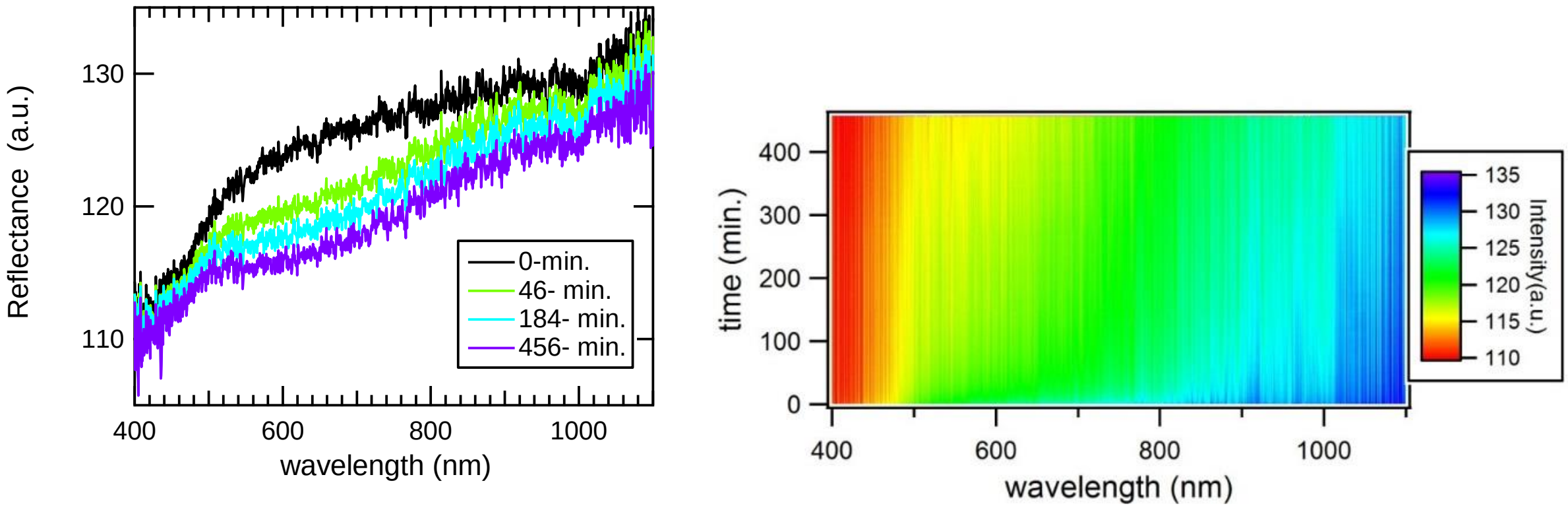

Figure S14: Reflectance spectra during measurement 2 (7 h) at given times (left) and as heatmap (right). The spectra clearly demonstrate the degradation of the sample over time.

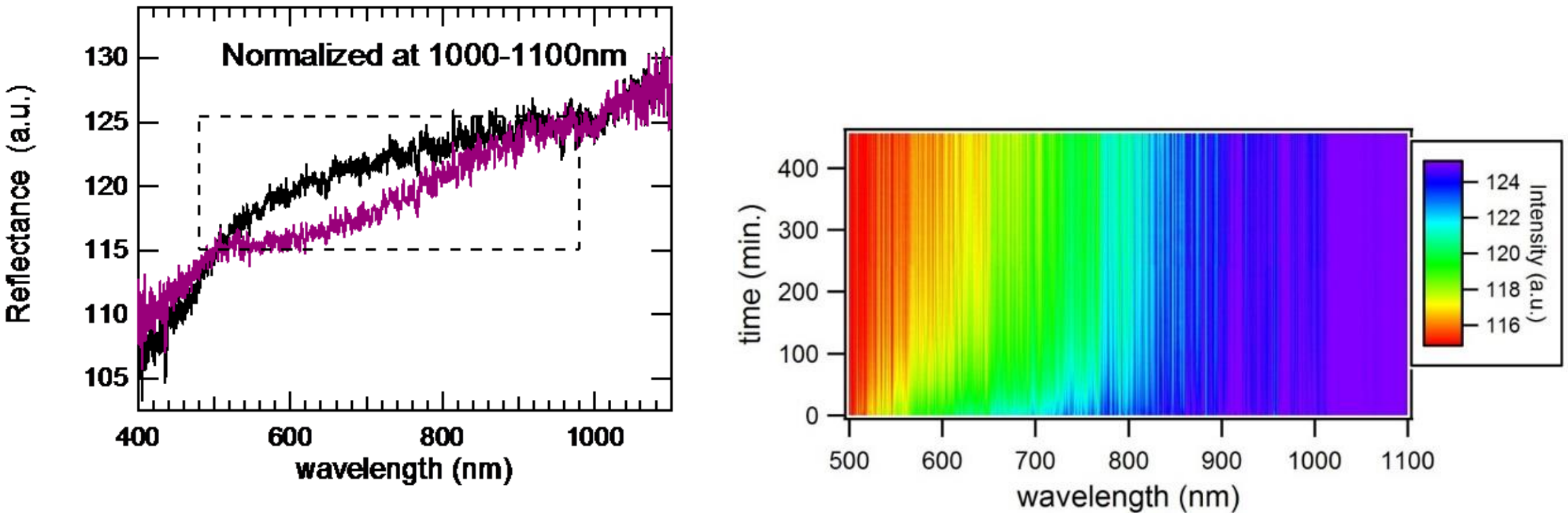

Figure S15: Reflectance spectra during measurement 2 (7 h) at given times (left) and as heatmap (right) normalised to the region 1000-1100 nm. Further to the overall loss in intensity, this demonstrates a stronger decrease in the region 500-900 nm than in the overall spectra.

SI References

[S1] V. Petříček, M. Dušek, L. Palatinus, *Z. Kristallogr. Cryst. Mater.* **2014**, *229*, 345.

[S2] K. Brandenburg, H. Putz, 2005. DIAMOND Version 3. Crystal Impact GbR, Postfach 1251, D-53002 Bonn, Germany.

[S3] W. Kraus, G. Nolze, *J. Appl. Crystallogr.* **1996**, *29*, 301.

[S4] A. Franz, A. Hoser, *Journal of large-scale research facilities* **2017**, *3*, A103

[S5] a) [1] X. Gonze, F. Jollet, F. Abreu Araujo, D. Adams, B. Amadon, T. Applencourt, C. Audouze, J.-M. Beuken, J. Bieder, A. Bokhanchuk, E. Bousquet, F. Bruneval, D. Caliste, M. Coe, F. Dahm, F. Da Pieve, M. Delaveau, M. Di Gennaro, B. Dorado, C. Espejo, G. Geneste, L. Genovese, A. Gerossier, M. Giantomassi, Y. Gillet, D.R. Hamann, L. He, G. Jomard, J. Laflamme Janssen, S. Le Roux, A. Levitt, A. Lherbier, F. Liu, I. Lukacevic, A. Martin, C. Martins, M.J.T. Oliveira, S. Ponce, Y. Pouillon, T. Rangel, G.-M. Rignanese, A.H. Romero, B. Rousseau, O. Rubel, A.A. Shukri, M. Stankovski, M. Torrent, M.J. Van Setten, B. Van Troeye, M.J. Verstraete, D. Waroquiers, J. Wiktor, B. Xu, A. Zhou, J.W. Zwanziger, *Comput. Phys. Commun.* **2016**, *205*, 106; b) [1] X. Gonze, J.-M. Beuken, R. Caracas, F. Detraux, M. Fuchs, G.-M. Rignanese, L. Sindic, M. Verstraete, G. Zerah, F. Jollet, M. Torrent, A. Roy, M. Mikami, P. Phosez, J.-Y. Raty, D. C. Allan, *Comput. Mater. Sci.* **2002**, *25*, 478–492.

[S6] a) J. P. Perdew, K. Burke, M. Ernzerhof, *Phys. Rev. Lett.* **1996**, *77*, 3865; b) J. P. Perdew, K. Burke, M. Ernzerhof, *Phys. Rev. Lett.* **1997**, *78*, 1396.

[S7] D. R. Hamann, *Phys. Rev. B* **2013**, *88*, 085117.

[S8] T. Björkman, *Computer Phys. Commun.* **2011**, *182*, 1183.